# Preliminary Evaluation of Interactive Search Engine Interface for Visually Impaired Users

Aboubakr Aqle[1], Kamran Khowaja[2], and Dena Al-Thani[3]
[1, 2, 3]Information & Computing Technology, College of Science & Engineering, Hamad Bin Khalifa University. Qatar

Corresponding author: Aboubakr Aqle (e-mail: aaqle@mail.hbku.edu.qa).

This work was supported by QNRF-GSRA and Mada assistive technology center in Qatar.

**ABSTRACT** This work designs, evaluates, and improves a proposed search engine interface for Visually Impaired (VI) users to efficiently perform web search activities. Our conceptual modeling technique is based on Formal Concept Analysis (FCA) that is used for data analysis. This approach highlights the hierarchized approach to represent the discovered concepts. It is combined with context interactive navigation in an interface which is called interactive search engine (InteractSE). This interface aims to reduce the time and effort required by the VI users to browse search results. InteractSE was evaluated by experts using Nielsen's heuristics and Web Content Accessibility Guidelines (WCAG) 2.0 for its usability and accessibility. The analysis was carried out based on the usability problems identified and their average severity ratings. The results show that the most frequently violated heuristics from Nielsen's set are consistency, documentation, and the average severity rating of all the problems is minor. The results also show that the most frequently violated WCAG 2.0 guidelines are distinguishable, followed by navigable and affordance. The average severity rating of all the problems found using WCAG 2.0 guidelines is also minor. The results show that Nielsen's heuristics and WCAG 2.0 guidelines contributed to identifying several usability problems, which might have missed out if either of them was used alone.

## I. INTRODUCTION

The web has been a blessing for people with visual impairment (VI) by allowing them to access a huge amount of information that was previously unobtainable via braille or audio interpretations. Since the beginning of the previous decade and with the development of screen readers, VI users are having instant and limitless access to information. This, in return, has supported their independence and integration in workplaces and educational settings.

However, despite this advancement, web pages are becoming more and more complex for a screen reader to access. Thus, VI web surfers are left with many challenges hindering their interaction [1, 49] among which performing a web search task that can be very challenging [2, 48].

The search engine results page (SERP) is a part of the web search task that contains the title, URL address and snippet for describing the web page for each result. SERP may contain other ads that cause delay for the VI users because of the screen reader's linear approach. These challenges define the VI needs for a new web search interface that accelerates the searching process. At the same time, the proposed interface should be tested with the VI for its usability.

Studies in the field have long stressed the fact that accessibility cannot substitute usability; this highlights that accessibility and usability must be considered [3-5]. Studies such as Correani et al. [6] and Hudson [3] have shown that websites can conform to the accessibility guidelines with many usability issues that hinder the users' interaction remaining.

There is no clear understanding between the relationship between accessibility and usability despite being discussed by several researchers such as [7] as cited in [8]. These authors discussed three views on the relationship between accessibility and usability: 1) people with and without a disability are different; thus the usability problems they experience also differ [7]. 2) Usability problems may include accessibility problems as well [9]. 3) The term "universal usability" covers both types of problems, i.e. usability as well as accessibility [5]. This indicates that the concept of typical usability can be expanded by including the experiences of disabled people, and there is a need to evaluate all the interfaces.

There is no clear, widely accepted definition of usability that is applied in practice, because it is not a property that can be measured using a measuring instrument like a thermometer is used to measure the body temperature of a human being [38-40] as cited in [41].

It is important to fix usability problems sooner (during the early design) rather than later (once a prototype is ready for the end user.) The fixation of usability problem during the later stage will cause more cost than getting them fixed as early as possible [50]. Usability testing can be conducted using formative or summative methods. The formative methods are typically used to identify specific usability problem, while, the general usability assessments are conducted with summative usability methods [41, 42]. Each method can be in the form of user-based evaluation or expert-based evaluation. A user-based evaluation uses a set of users who are given a set of representative tasks to be performed on the application/system/interface, while an expert-based evaluation is a structured inspection of an application, system or interface by one or more experts [51]. The evaluation relies on the practical and theoretical skills of the experts, and these skills allow them to perform a set of tasks based on a given set of guidelines or standards [52]. This evaluation is typically performed before user-based evaluation as experts can pinpoint obvious flaws that need to be fixed [51].

Ivory and Hearst [43] and Nielsen [44] have identified a number of formative methods that uses expert-based evaluation; these methods include guideline review, cognitive walkthrough [11, 12], pluralistic walkthrough, heuristic evaluation [15], perspective-based inspection, feature inspection, formal usability inspection, consistency inspection, and standards inspection. They have also identified methods that involve users and testers, and classified these methods into five classes, namely, testing, inspection, inquiry, analytical modeling, and simulation. Dix et al. [10] has additionally described three different approaches to evaluating the system through experts including Goals, operators, methods and selection (GOMS) [13], Keystroke-level model [14], and use of previous results as a basis to prove or disprove different aspects of the design.

Lewis has highlighted that "Before conducting testing with users, part of the preparation of a study should include an inspection method such as heuristic or expert evaluation [42, p.665]." Since the users are involved at the later stage of product development, experts are used to evaluate a product during its early stage of development. The experts have domain-specific knowledge; they can evaluate the product and identify the usability problems that need to be fixed before actual users start using it.

Among the various usability evaluation methods that uses experts as a part of evaluation study, researchers frequently use heuristic evaluation to evaluate a product because it is cheap, intuitive and easy to motivate experts to evaluate, requires no advance planning, and is used in the early development process [15]. One advantage of these heuristics is that they can be modified and expanded to fulfil the needs of specialized domains. The review of literature has shown that a number of specialized set of heuristics have been created to identify usability problems from the perspective of a specific domain. These specialized set of heuristics have been created for ambient displays [16], collaborative tasks [17], human-robot interaction [18], persuasive health technologies [19], video games [20], e-learning applications for children [21], deaf web user experience [22], and interactive systems for children with autism spectrum disorder (ASD) [23] among others. No specialized heuristics have been developed to evaluate a website for visually impaired people. To the best of our knowledge, there is no specialized set of heuristic that can be used in this research to evaluate search applications developed for visually impaired people. Thus, a set of heuristics by Nielsen [24] are used in this research in addition to the Web Content Accessibility Guidelines (WCAG) guidelines 2.0 to evaluate search engine being developed as a part of ongoing research, in terms of usability and accessibility.

The aim of this research is to conduct a usability evaluation and improve the interface for visually impaired people to search and browse results with experts using WCAG 2.0 guidelines and Nilsen's set of heuristics. Section II presents the related work on the topic, the accessible search engine design described in section III, and the study design is described in section IV. The results are presented in section V. Section VI presents the interface enhancement, while the last section presents the conclusion.

## II. RELATED WORK

### A. ACCESSIBLE SEARCH ENGINE

Online information seeking has become one of the most frequently engaged tasks that people carry out in their daily lives. VI users use speech-based screen readers to access search engines. Given that only text is rendered in a serial nature, the VI web surfer perceives the web very differently than their sighted peers. Hence, their performance is different as studies have shown [2, 25-27]. These studies investigated the differences and highlighted the challenges occurring during such activity. They emphasized that the result in the exploration stage, where the user skims through the set of search results, is the most challenging and time-consuming. This is not surprising given the issues the VI users face on the web. Ivory and Chevalier [25] and [27] concluded that VI users spent more than double the time sighted users spend when examining a search results page. This, in turn, affected their overall performance and integration in workplaces and educational teams [28].

Even though this issue has long been highlighted in web accessibility research, very few have attempt to address it. Parente [29] was one of the very early attempts to address this issue. Influenced by Marchionini et al.'s [30] Agileviews framework, Parente developed and evaluated the audio enriched links which present the user with a speech-based summary of a webpage. The summary consists of the webpage title, number of headers, and other content related statistics that can give the user an overview of the content of the page.

Sahib et al. [2] highlighted a number of challenges which the VI web surfer encountered when searching the web and described result exploration as the most problematic. As a result of this study, Sahib and her research group introduced an integrated tool that allows VI users to keep track of search progress and manage search results [31]. Such a tool will allow the user to save search results while going through a large set of search results. The user can easily return to the search results of interest. This feature seems to support the user in this stage. When evaluating the tool with VI participants, the participants were highly satisfied with the usage of the features, which they refer to as a seamless and easy way to handle search results within the tool. In this paper, we attempt to tackle this problem via an algorithmic approach, which is introduced in section III.

### B. ACCESSIBLE INTERFACE EVALUATION
*1) Web Content Accessibility Guidelines (WCAG)*

The W3C- WAI WCAG guidelines are the most renowned web accessibility guidelines. Starting in the year 2000, the WAI produced a number of guidelines to help address accessibility, the most popular of which is the WCAG as it aims to address the accessibility of web pages and make web interaction available for all.

The first set of guidelines, WCAG 1.0, was released in 1999. This set of guidelines mainly catered for the accessibility issues that occurred in static web pages. To cater to the needs of web 2.0, in the year 2000, the WAI group started planning for a newer version. A draft of this newer version, WCAG 2.0, was announced in 2003. WCAG 2.0 had four primary principles comprising perceivable, operable, understandable and robust:

• Perceivable: the content presented must be apparent and clear to a diverse set of users at all time.

• Operable: the web component must be operable using a variety of means. This encourages the web developers and designers to think of different ways of interaction to cater for the different modes of interaction.

• Understandable: the content must be understandable to all.

• Robust: the content should be rendered using different assistive technology applications in a seamless and efficient way.

For each principle, there is a set of guidelines that need to be adhered [32]. Each guideline is supplemented by success criteria to help web developers and experts when checking conformance of the guidelines. The web developers or experts then rank the conformance of the guidelines using the levels A, AA, or AAA, where level A is considered the minimum conformance level [33].

The guidelines included in each principle are as follow. One word of each guideline is written in a square bracket. These words are the shorter names of the guidelines and will be referred to in the later sections.

The four guidelines in the principle called "Perceivable" include:
• Text [Alternatives]
• Time-based [Media]
• [Adaptable]
• [Distinguishable]

The five guidelines in the principle called "Operable" include:
• [Keyboard] Accessible
• [Enough Time]
• Seizures and Physical [Reactions]
• [Navigable]
• Input [Modalities]

The three guidelines in the principle called "Understandable" include:
• [Readable]
• [Predictable]
• [Input Assistance]

A guideline in the principle called "Robust" include:
• [Compatible]

Conformance to WCAG 2.0 is the most used web accessibility evaluation method [34]. This evaluation can either be done automatically using a software tool or manually by an expert. There are several automated tools, some of which the WCAG 2.0 recommends. However, the research field has long criticized this approach by stressing that the outcome of using such a tool is not reliable and human intervention in such practices is an absolute necessity [25, 34]. Therefore, WCAG 2.0 also suggested that websites could be checked manually for their conformance to the WCAG guidelines. Such a process is called guidelines review. The process includes one or more evaluators to check manually whether a website satisfies the set of guidelines and their success criteria.

*2) Nielsen's set of heuristics*

Nielsen and Molich [15] developed an initial set of principles referred to as heuristics to inspect if all the elements present in the interface follow the principles. These heuristics (principles) are broad rules of thumb than a specific set of usability guidelines to follow. The initial set included nine heuristics. Later, Nielsen came up with a set of ten heuristics [24] based on the work at an individual level. These ten heuristics are as follows. One word of each heuristic is written in a square bracket. These words are the shorter names of the heuristics and will be referred to in the later sections.
1. [Visibility] of system status
2. [Match] between system and the real world
3. [Consistency] and standards
4. [Recognition] rather than recall
5. Aesthetic and [minimalist] design
6. User [control] and freedom
7. [Error] prevention
8. [Flexibility] and efficiency of use
9. Help users recognize, diagnose, and [recover] from errors
10. Help and [documentation]

## III. ACCESSIBLE SEARCH ENGINE DESIGN

This section discusses the proposed design of the web search interface for the visually impaired users called interactive search engine (InteractSE).

InteractSE is a Google search interface targeting visually impaired users that minimize the representation text of the search results that need to be read by the screen reader. It allows the user to have an overview of the target web page before navigating to it.

Web scraping [47] is a powerful technique for data extraction from the World Wide Web (WWW); in our case scraping is used to extract data from SERP. After scraping the search engine for the required search query, Google search results are pre-processed by Natural Language Processing (NLP) stage to exclude the stop-words, get the root keywords by the stemmer, and make results ready for the discovery of the concept by Formal Concept Analysis (FCA) process. Concepts are the base component of human thinking, reasoning and FCA [35]. FCA is a clustering method for knowledge representation that covers the maximum number of documents sharing for a maximum number of attributes [36]. The final stage is the results presented in a multilevel tree structure of the discovered concepts in a hierarchical order.

The design workflow of the web search interface enables the visually impaired users to narrow the search results by selecting the main keyword at the tree level known as a concept. We can notice from Figure 1 that the user received nine results at the first level of the tree, then seven results at the second level, and three results at the third level of the tree. The user navigates the tree using the down arrow key only what changes the results at the list area.

The interface has three parts: 1) Query input field. 2) Search results represented in multilevel tree of the keywords. 3) List of the search results of the selected keyword that match the hierarchy of the tree, as shown in Figure 1.

## IV. STUDY DESIGN

### A. PARTICIPANT AND RECRUITMENT

Nielsen and Molich [15] suggested recruiting about five experts (with at least three) as they are able to identify more than 75% of the usability problems. Thus, five experts are recruited for this research.

The experts who were chosen for the study involve research and academic university staff who conduct research and evaluation in HCI or interface design experience. They have worked in web design and have the required experience for the heuristic evaluation. The invitations were sent to five experts who confirmed their participation in this study, and completed the experiment with their feedback.

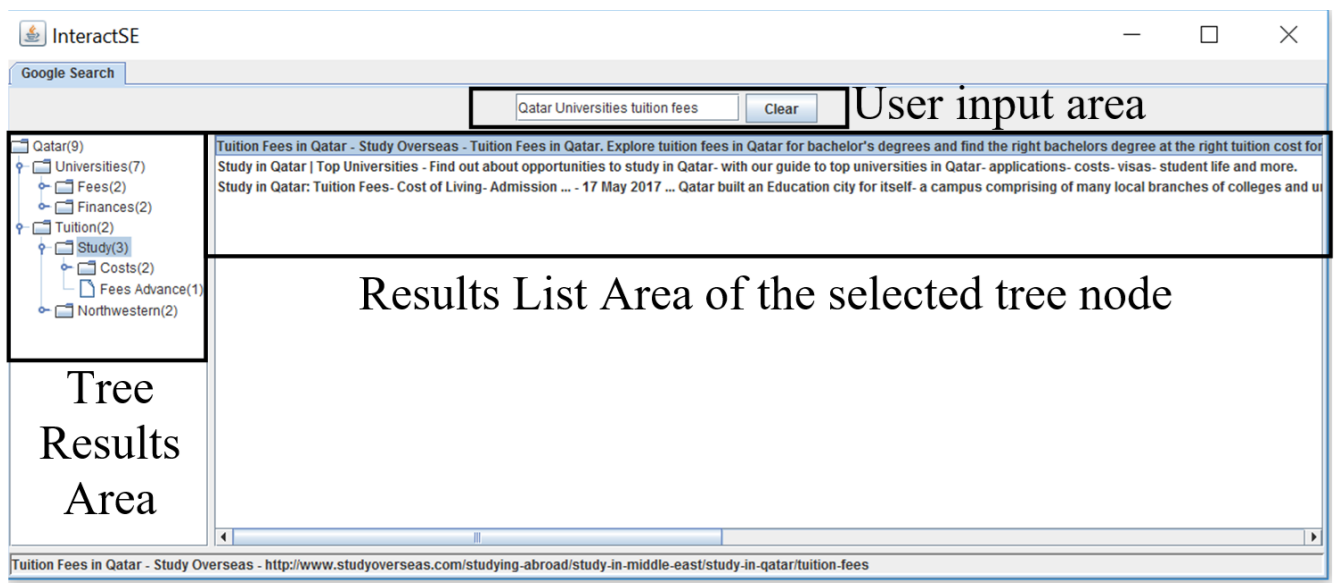

**FIGURE 1.** Web Search Interface Design

Since the most famous screen readers used are: 1) Job Access With Speech JAWS. 2) Non-Visual Desktop Access NVDA, the expert participants were asked about their experience with these tools as shown in Table I for their demographic profiles.

Table I: PARTICIPANTS' DEMOGRAPHIC PROFILES

| Characteristic | Values |
| --- | --- |
| Age | Average 31 years of 3 (35), 2 (25) |
| Gender | 3 Female, 2 Male |
| Education | 3 Master, 2 Doctorate Degree |
| Occupation | 4 Employed, 1 PhD Student |
| HCI Course(s) | 4 Yes, 1 No |
| HCI Experience | 3 Expert, 2 Advanced |
| Screen Reader | 1 JAWS, 1 NVDA, 3 Never Used |

### B. INSTRUMENTS USED

The main instrument used in this study is the usability reported problems with the severity ratings against the heuristic of Nielsen's set and WCAG 2.0 guidelines. For each usability problem, expert was asked to mention the usability heuristic or accessibility guideline broken, describe the usability problem in detail, potential solution from their own perspective, and the severity rating of the usability problem. The severity rating of the usability problem can be from 0 through 4, where 0 means I do not agree that this is a usability problem at all, 1 means cosmetic problem only that need not be fixed unless extra time is available on project, 2 means minor usability problem and fixing this should be given low priority, 3 means minor usability problem and important to fix with high priority, and 4 means usability catastrophic that imperative to fix this before product can be released. Expert feedback is important to have a better understanding of their views and evaluation of the designed interactive web search interface, InteractSE, for the visually impaired users.

Lewis has highlighted four most widely used standardized usability questionnaires to assess the perception of usability at the end of a study [42]. These questionnaires are also cited in national and international bodies like American National Standards Institute, and International Organization for Standardization. The questionnaires include Questionnaire for User Interaction Satisfaction, Software Usability Measurement Inventory, Post-Study System Usability Questionnaire (PSSUQ), and Software Usability Scale (SUS).

As cited in [42], the SUS by Brooke [45] is probably the most widely used standardized usability questionnaire [46] containing ten questions with alternating positive and negative tones, as shown in Figure 2. Considering the wide applicability of the questionnaire, it is also used in this study.

FIGURE 2. System Usability Scale

The System Usability Scale covers the following usability measurement:

- Effectiveness: users' ability for completing the tasks by the system with output quality.
- Efficiency: level of consumed resource in performing the tasks.
- Satisfaction: users' subjective responses to using the system.

SUS cover different forms of the system usability, like the complexity and need for training or support, and thus can be considered a high measuring unit for validating the usability of a system.

### C. STUDY PROTOCOL

The below scenario was carried out for the study protocol:

1. Participants were invited to the designed interface evaluation experiment by email with Nielsen's set and WCAG 2.0 guidelines in the attachment, and they confirmed their acceptance.

2. The experiment was performed with the participants individually face-to-face at the campus research complex. At the beginning of each session, the interface design was explained in detail and training was given to the participant explaining how to use and search the web using the system for their query search input. The participants were informed of the purpose of the system and their evaluation to highlight the usability problems they faced during the experiment.

3. Participants were given the heuristic set and guidelines to be used as a reference during their exploration of the system. We provided guidelines for the web accessibility evaluation's tools of the screen readers: JAWS and NVDA for the participants, to have a better understanding of how to use these tools.

4. Participants were asked to write during their evaluation the number of the broken heuristic or guideline, problem description in brief, their recommendation to overcome this broken heuristic or guideline, with severity ratings between 0 and 4. Severity rating with 0 assigned to 'not a problem', one to 'cosmetic problem only', two to 'minor', three to 'major' and four to 'usability catastrophe'.

5. After the exploration and system evaluation, the System Usability Scale form was given to the participants to describe their opinion for each statement of the ten points about the system. The criteria for rating each point was based on a scale of 1 (strongly disagree) through 5 (strongly agree).

### D. DATA ANALYSIS

The first analysis is based on the following two parameters. Both parameters are separately calculated for each heuristic of Nielsen's set and WCAG 2.0 guidelines.

1. Number of usability problems identified: It is calculated as a sum of all the problems identified by the experts for each Nielsen's set or WCAG 2.0 guidelines.
2. Average severity ratings: The average severity rating is calculated for all the problems identified by the experts using Nielsen's set or WCAG 2.0 guidelines.

The second analysis is based on the System Usability Scale which is a simple ten-item attitude Likert scale that is giving a global view of subjective assessments of the system's usability.

## V. RESULTS

The results of evaluations using Nielsen's set of heuristics and WCAG 2.0 guidelines are discussed in the following sub-sections. The results are presented based on the number of usability problems found and the average severity ratings.

### A. Nielsen's Set of Heuristic
#### 1) Number of usability problems found:

Figure 3 shows the usability problems and the average severity ratings of all the usability problems identified in each heuristic by Nielsen [24]. The left vertical axis represents the number of usability problems, while the right vertical axis represents the average severity ratings of all the usability problems. Each stacked column represents one of Nielsen's heuristic and shows the number of usability problems identified for one or more of the four severity ratings (cosmetic, minor, major or catastrophe). The line that runs through the markers shows the average severity ratings of all the usability problems.

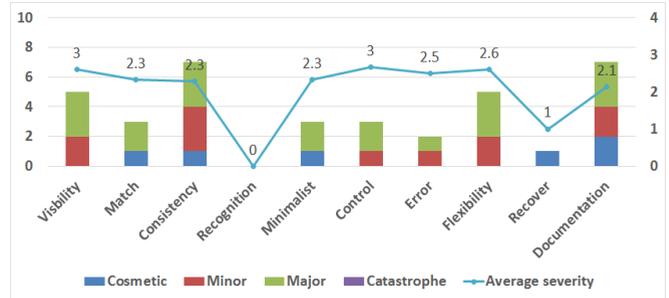

**FIGURE 3.** Usability problems identified using Nielsen

The most commonly broken heuristics are consistency and documentation (each has N=7) followed by visibility and flexibility (each has N=5). Some comments of the experts for the most frequently violated/broken heuristics are given in Table II. The first-five comments are related to the consistency heuristic, while the remaining comments are related to the documentation heuristic.

#### 2) Average severity ratings:

The average severity ratings of all the problems show that they are minor. This shows that it may have some impact on usability. Therefore, it is better to fix them.

### B. WCAG 2.0 Guidelines

Figure 4 shows the usability problems identified and the average severity ratings of all the usability identified in each guideline of WCAG 2.0. The information and its format presented at the vertical axes in Figure 4 are the same as Figure 3.

Table II: EXPERTS' COMMENTS BASED ON NIELSEN'S HEURISTICS

| Heuristic broken | Problem Description | Recommendation | Severity Rating |
|---|---|---|---|
| Consistency | The system has F1 and F12 keys reserved for the data collection and analysis of search results. | The use of F1 is typically reserved to start and navigate the help. This contradicts with the standardized key. Use other keys for the shortcuts. | 1 |
| Consistency | Pressing an enter key at the search text field doesn't do anything. | Pressing an enter key should initiate the search operation as done in typical search engines. | 2 |
| Documentation | There is no help in the system. | The system should provide help on the use of the system. | 3 |
| Control | It is hard to know which region I am in. | The screen reader could probably mention the region. | 3 |
| Flexibility | The system does not provide any shortcut keys | The system should provide shortcut keys to perform all the operations supported by the system as well as to ease in the navigation. | 2 |
| Error | Clearing the search query by clicking on the clear button is only clearing the text field but not the results | It should clear the tree and list view as well. | 2 |
| | presented in the tree and list. | | |
| Match | User cannot perform another search before clearing the text of first query | Treat each text as a new query. There is no need to toggle the caption of the button. This creates a confusion. | 3 |
| | | For instance, the "HBKU" in the description should be pronounced as "Hamad Bin Khalifa University." | |

Each expert was asked to classify the identified usability problem into 1 of 61 success criterion. However, due to limited space, the number of these problems are grouped and shown based on the guideline with which they are associated. It is to be noted that as per the WCAG 2.0, not all 12 guidelines are testable at their own but their corresponding success criterion is testable.

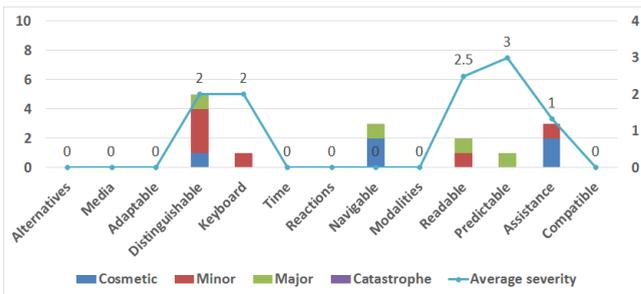

**FIGURE 4.** Usability problems identified using WCAG 2.0

*1) Number of usability problems found:*

The most commonly broken guidelines are the distinguishable (N=6) followed by navigable and affordance (each has N=5). Some of the experts' comments for the most frequently violated guideline are given in Table III.

*2) Average severity ratings:*

The average severity ratings of all the problems identified show that they are minor.

Table III: EXPERTS' COMMENTS BASED ON WCAG 2.0 GUIDELINES

| Guideline broken | Problem Description | Recommendation | Severity Rating |
|---|---|---|---|
| Distinguishable | The minimum font size should 12 as per the upcoming guidelines of WCAG. | It will be good to keep the minimum font size of text to 12. | 1 |
| Distinguishable | There is no line space between two items of the list. | Give the recommended line space for the ease in navigation. | 2 |
| Keyboard | Keyboard shortcuts non-conformant to standards | Fix to conform to standards | 2 |
| Assistance | no error suggestions | tell the user when clicked on search, tell them what to do next | 1 |
| Navigable | A bit confused with the focus during the navigating the regions | A alt text reading the region name | 1 |
| Readable | The full form of the abbreviations are not pronounced | As per the guideline, the user should be informed about the full form followed by the acronym. | 3 |

Table IV shows the number of usability problems identified and its percentage using Nielsen's set of heuristics referred to as "NE" in the table and WCAG 2.0 based on the severity as well as the sets (NE and WCAG).

Table IV: COMPARISON OF USABILITY PROBLEMS IDENTIFIED USING NIELSEN AND WCAG 2.0

| | | System | | Total Issues |
|---|---|---|---|---|
| | | NE | WCAG 2.0 | |
| Severity Rating (SR) | Count of SR 4 | 0 | 0 | 0 |
| | % within Severity | 0.00 | 0.00 | 0.00 |
| | % within Set | 0.00 | 0.00 | 0.00 |
| | Count of SR 3 | 19 | 4 | 23 |
| | % within Severity | 82.61 | 17.39 | 100.00 |
| | % within Set | 52.78 | 26.67 | 45.10 |
| | Count of SR 2 | 11 | 6 | 17 |
| | % within Severity | 64.71 | 35.29 | 100.00 |
| | % within Set | 30.56 | 40.00 | 33.33 |
| | Count of SR 1 | 6 | 5 | 11 |
| | % within Severity | 54.55 | 45.45 | 100.00 |
| | % within Set | 16.67 | 33.33 | 21.57 |
| Total | Total count | 36 | 15 | 51 |
| | % covered | 70.59 | 29.41 | 100.00 |

It can be seen that (N=36, 71%) of the total problems have been identified using Nielsen's set of heuristics, while the remaining (N=15, 29%) of the total problems have been identified using WCAG 2.0.

No problem was identified for the catastrophe severity. Further, almost half of the problems (N=23, 45%) were identified as major, followed by minor and cosmetic. Based on the problems identified within the set of NE, slightly more than half (N=19, 53%) of the problems were identified as major, followed by minor and cosmetic. While, based on the problems identified within the set of WCAG 2.0, the number and percentage of the problems identified across the three severity levels, i.e. major, minor and cosmetic, are the same.

Based on the problems identified within the set of NE, it can be seen that slightly more than half (N=19, 53%) of the problems were identified as major, followed by minor and cosmetic. While, based on the problems identified within the set of WCAG 2.0, it can be seen that (N=6, 40%) of the problems were identified as minor, followed by cosmetic and major. There is a subtle difference between the numbers of problems found across the severity ratings.

### C. SYSTEM USABILITY SCALE (SUS)

SUS was used to evaluate the usability of the designed web application InteractSE. The evaluation and calculation were calculated based on SUS guidelines [37]. The result of the survey's questions was computed using the calculation rule of SUS and the mean of the five participants is presented at Figure 5.

The value of SUS scores is distributed between 60 and 100 with the smallest value falling in 60's and the largest value falling in 100.

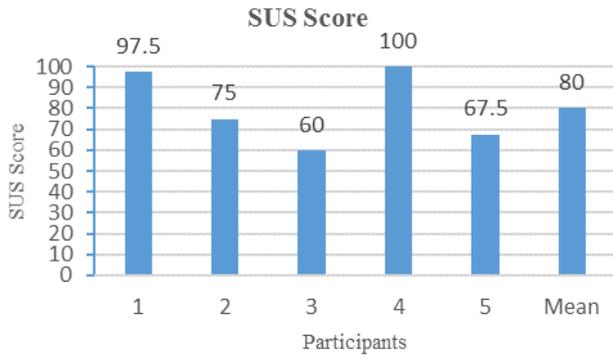

**FIGURE 5.** SUS Score Results and Mean

The average of SUS score for InteractSE was 80 out of 100. Considering a benchmark of 68 defining a categorization of average and a threshold of 72 required for a good usability rating, as shown in Figure 6 [37]. The result for this study obtains a usability rating of good. Hence, the designed web needs a minor improvement and enhancement before it is used by the public.

## VI. INTERFACE ENHANCEMENTS

Many enhancements are achieved in the web search interface. Web page title and summary description at the list component were extended on multiple lines and the horizontal scrollbar removed. The spacebar was placed between the items at the list component as a separator. Font size was adjusted to the window size, to be changed automatically to be smaller or larger based on the window's aspect ratio. The enter key was defined as an active key to start the search process as the search button click action.

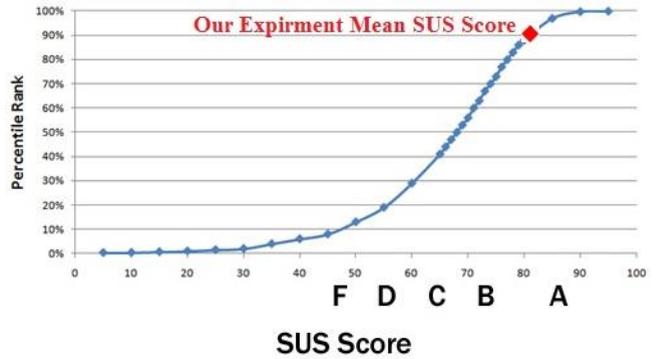

**FIGURE 6.** SUS Threshold and our Experiment Mean Result

Help was added to the interface to assist the end-user, and the default shortcut key F1 was assigned to the help function. The shortcut key Alt+W was assigned to the "Where I am" function for the end-user to be aware of the cursor's location standing at which region of the interface. All these changes can be noticed in Figure 7 for the new enhanced interface. All updated and new shortcut keys are summarized in Table V.

Table V: UPDATED / NEW SHORTCUT KEYS

| Shortcut Key | Action |
|---|---|
| F1 | Help |
| F12 | Terminate Experiment & Collect Data |
| Alt+W | Where I am |
| Ctrl+W or Ctrl+F4 | Close Tap Page |

## VII. CONCLUSION

This work proposed a new web search interface targeting visually impaired VI users. The proposal interface is based on discovering concepts through Formal Concept Analysis. VI users interact with the interface to collect concepts as keywords that narrow the search results to get the target web pages containing the required information with the minimum effort and time required.

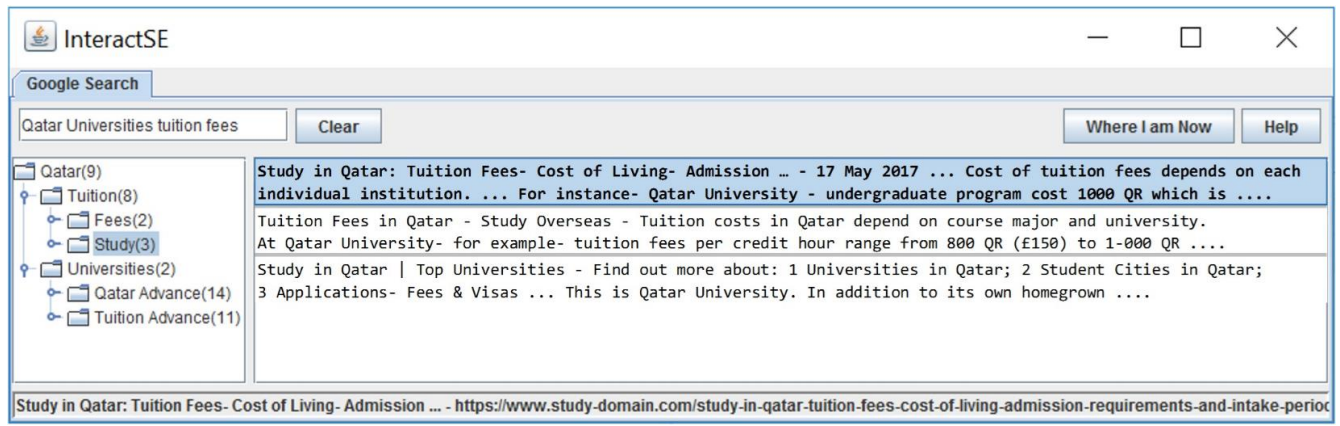

**FIGURE 7.** Interface Enhancements

This research presents a usability evaluation of the search interface that is developed for the VI users. The usability evaluation was carried out with experts in the field of HCI and accessibility using a set of heuristics by Nielsen and a set of WCAG 2.0 guidelines.

Both sets contributed to identifying a number of usability problems based on the details mentioned in the description of each heuristic and an individual guideline.

While following the guidelines of WCAG 2.0, one can ensure that an application (standalone or web-based) is accessible by everyone including a person with disability/impairment. On the contrary, following Nielsen's heuristics or any other user interface guideline (like eight golden rules of interface design by Shneiderman), one can ensure that usability problems have been fixed before anyone including the person with any disability/impairment starts using that application.

Both, the Nielsen's heuristics and WCAG 2.0 guidelines have a different purpose. They cannot be preferred over one another but can complement each other. This has been seen in the usability evaluations conducted in this research.

Although the application had limited functionalities, for instance, having no videos or images, WCAG 2.0 contributed to finding a number of usability problems that had otherwise gone unnoticed with NE. An application with more features and functionalities may reveal more usability problems from the perspective of WCAG 2.0 than NE. This requires further investigation. In the future, researchers can evaluate multiple applications using both NE and WCAG 2.0. Researchers can also develop a set of guidelines by making use of WCAG 2.0 guidelines, Nielsen's heuristics and the web-based guidelines to evaluate websites for visually impaired people.

## ACKNOWLEDGMENT

This contribution was made possible by GSRA grant No. 04-1-0514-17066 from the Qatar National Research Fund (a member of Qatar Foundation). The statements made herein are solely the responsibility of the authors. We gratefully acknowledge Mada assistive technology center in Qatar support in this project. Mada's support has had a huge impact on the quality of work produced in this project.

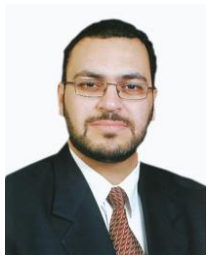

**ABOUBAKR AQLE** is currently a PhD Candidate of Computer Science and Engineering at Hamad Bin Khalifa University in Doha, Qatar. Aqle has BSc major in Computer Science from Qatar University at 2002, and MSc in Computer Science and Engineering from Qatar University at 2015. His master thesis was based-on real-time systems analysis for the hidden web databases. Before undertaking doctoral studies in 2016 at Hamad Bin Khalifa University, Aqle worked for different techno-functional and managerial ICT positions for more than 14 years in multi-sectors of private, government and semi-government institutions. He started his career as Programmer, then as System Analyst, after that as Implementation Team Leader to a Project Manager, and finally as ICT Solutions Manager. He is actively involved in Qatar University research project of "Analytics based Interface Transformation for Web Databases" for National Priorities Research Program (NPRP) that is a funding program for Qatar National Research Fund (QNRF). He applied new information extraction methodology that is domain independent and gave promising results. His work has appeared in several of international conferences and journals. Aqle assisting many undergraduate and master students at Qatar University for the Mobile App development framework and Formal Concept Analysis approach for the hidden web data analytics and model representation. His research topics of interest are Formal Concept Analysis, Hidden Web Data Analytics, Concepts Extraction and Browsing, Semantic and Structural Analysis in documents, and Text Summarization Techniques.

**DENA AL-THANI** In 2009, Dr. Dena Al Thani graduated with an MSc in Software Engineering with distinction from the University of London. Her thesis in Human Computer Interaction (HCI) investigated visually impaired and sighted people's collaborative computer use and proposed technical approaches to support it. Following her graduation she managed the online portal and integration platforms for Ooredoo-Qatar before joining Queen Mary University of London's PhD studentship program to continue research in her field of interest. In 2016, Dr. Dena Al Thani was awarded her PhD in Computer Science, her thesis was titled "Understanding and Supporting Cross-modal Information Seeking". The thesis explored the under-investigated area of cross-modal interaction and inclusive design and evaluation. Her academic and research vocation is to explore and demonstrate how HCI as a field of applied enquiry can contribute to building a more inclusive society. In addition to her research work at Queen Mary University of London, Dr. Dena has worked as a teaching assistant in three Computer Science modules including database systems and programming. She has obtained a postgraduate certificate of Learning and Teaching in Higher Education, and is an associate member of the Higher Education Academy in the UK.

**KAMRAN KHOWAJA** Kamran Khowaja received his BCS and MCS from Isra University, Pakistan, MSc from Asian Institute of Technology (AIT), Thailand and PhD from the University of Malaya (UM), Malaysia. He is currently working as an Associate Professor at the Department of Computer Science, Isra University, Hyderabad, Pakistan before taking leave for postdoctoral research at Hamad Bin Khalifa University, Qatar. His areas of research include Human–Computer Interaction, Child–Computer Interaction and Serious Games.